\documentclass[letterpaper,10pt]{article} 
\usepackage{osameet3} %% use version 3 for proper copyright statement

\usepackage{amsmath,amssymb}
\usepackage[colorlinks=true,bookmarks=false,citecolor=blue,urlcolor=blue]{hyperref} %pdflatex
\usepackage{amsthm}
\usepackage{booktabs}
\usepackage{appendix}
\usepackage{pgfplots}
\usepackage{epstopdf}
\usepackage{graphicx}
\usepackage{amsfonts}
\usepackage{amssymb}
\usepackage{tikz}
\usetikzlibrary{shapes,arrows,backgrounds}
\usetikzlibrary{calc,fit,arrows}
\usepackage[font=small,skip=0pt]{caption}
\usepackage{subcaption}
\usepackage{adjustbox}
\usepackage{lineno}
\usepackage{setspace,wrapfig}
\usepackage{stfloats} % for positioning of figure* on the same page
\usepackage{array, multirow}
\newcolumntype{C}[1]{>{\centering\arraybackslash}p{#1}}
\usetikzlibrary{shapes}
\usepgfplotslibrary{groupplots}
\pgfplotsset{compat=1.13}
\usepgfplotslibrary{colormaps} 
\usetikzlibrary{pgfplots.colormaps} 
\usetikzlibrary[pgfplots.colormaps] 
\newtheorem{theorem}{Theorem}%[section]

\begin{document}

\title{Capacity Bounds for Optical WDM Channels with Peak Power Constraints}

\author{Viswanathan Ramachandran, Gabriele Liga, Astrid Barreiro, and Alex Alvarado}
\address{Eindhoven University of Technology, 5600 MB Eindhoven, the Netherlands.}
\email{v.ramachandran@tue.nl}
%%Uncomment the following line to override copyright year from the default current year.
\copyrightyear{2022}

\begin{abstract}
We investigate optical WDM transmission from the standpoint of an information-theoretic interference channel. Achievable rates that outperform treating interference as noise are presented and validated using split-step Fourier method simulations.
\end{abstract}
\section{Introduction}
\vspace{-2mm}
In optical transmission, data from different users are typically multiplexed into a single optical fiber using wavelength division multiplexing (WDM). The Kerr effect causes a signal in one wavelength to interfere with signals in other wavelengths, resulting in a stochastic nonlinear channel with memory when combined with chromatic dispersion. This channel is described by the nonlinear Schr{\"o}dinger equation (NLSE). Despite the multiuser nature of optical WDM channels, most papers in the literature resort to a single-user approach for their information-theoretic analysis. This usually involves the study of the capacity of a given user under different statistical assumptions for the interfering users~\cite{agrell2015influence}. Multi-user information theoretic studies exist, but only for highly simplified models \cite{taghavi2006multiuser,ghozlan2017models}. In \cite{ramachandran2022wdm}, we investigated the set of \emph{simultaneously achievable rates} for a realistic channel model for the first time, based on multi-user information theory. However, the channel model in \cite{ramachandran2022wdm} ignored the frequency separation between adjacent channels while modelling the strength of cross-channel interactions. Furthermore, the results in \cite{ramachandran2022wdm} were not validated via split-step Fourier (SSFM) simulations. The main contributions of this paper are: (i) a correction of the model in \cite{ramachandran2022wdm} to allow separation-dependent coupling between adjacent channels, and (ii) a validation of the results in \cite{ramachandran2022wdm} based on SSFM simulations. While the obtained results show improved lower bounds compared to treating interference as noise (TIN), the rates from simulations are smaller than that predicted by the model.
\vspace{-1mm}
\section{Channel Model}
\vspace{-2mm}
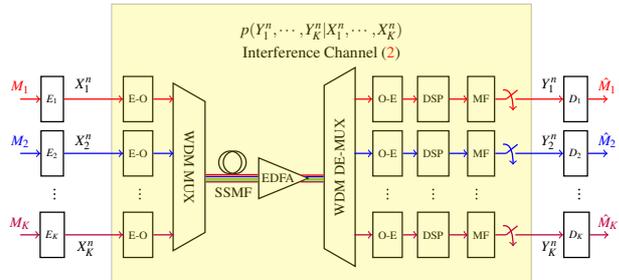
\begin{wrapfigure}{r}{0.6\textwidth}
\vspace{-2mm}
\centering
\scalebox{0.6}{\begin{tikzpicture}[thick]
\node (e1) at (-2.4,1.5) [rectangle, draw, minimum height=1.0cm]{\scriptsize ${E_1}$};
\node (e11) at (-2.4,0.3) [rectangle, draw, minimum height=1.0cm]{\scriptsize ${E_2}$};
\node (e2) at (-2.4,-1.5) [rectangle, draw, minimum height=1.0cm]{\scriptsize ${E_K}$};
\foreach \i in {-0.5,-0.6,-0.7} {\fill (-2.35,\i) circle(0.02cm);}
\foreach \i in {-0.5,-0.6,-0.7} {\fill (-0.48,\i) circle(0.02cm);}
\node (e3) at (-0.52,1.5) [rectangle, draw, minimum height=1.0cm,minimum width=0.28cm,align=center]{\scriptsize {E-O}}; 
\node (e33) at (-0.52,0.3) [rectangle, draw, minimum height=1.0cm,align=center]{\scriptsize {E-O}}; 
\node (e4) at (-0.52,-1.5) [rectangle, draw, minimum height=1.0cm,align=center]{\scriptsize {E-O}};
\node (e5) at (7,1.5) [rectangle, draw, minimum height=1.0cm,align=center]{\scriptsize {MF}};
\node (e55) at (7,0.3) [rectangle, draw, minimum height=1.0cm,align=center]{\scriptsize {MF}};
\node (e6) at (7,-1.5) [rectangle, draw, minimum height=1.0cm,align=center]{\scriptsize {MF}};
\node (e7) at (4.95,1.5) [rectangle, draw, minimum height=1.0cm,minimum width=0.05cm,align=center]{\scriptsize {O-E}};
\node (e77) at (4.95,0.3) [rectangle, draw, minimum height=1.0cm,minimum width=0.05cm,align=center]{\scriptsize {O-E}};
\node (e8) at (4.95,-1.5) [rectangle, draw, minimum height=1.0cm,minimum width=0.05cm,align=center]{\scriptsize {O-E}};
\node (e9) at (5.95,1.5) [rectangle, draw, minimum height=1.0cm,align=center]{\scriptsize {DSP}};
\node (e99) at (5.95,0.3) [rectangle, draw, minimum height=1.0cm,align=center]{\scriptsize {DSP}};
\node (e10) at (5.95,-1.5) [rectangle, draw, minimum height=1.0cm,align=center]{\scriptsize {DSP}};
\node[trapezium,draw,rotate=-90,minimum height=0.7cm,minimum width=0.4cm] (t1) at (0.6,0) {\small {WDM MUX}};
\node[trapezium,draw,rotate=90,minimum height=0.68cm,minimum width=0.25cm] (t2) at (3.90,0) {\small {WDM DE-MUX}};
\node (c1) at (1.6,0.1) [circle, draw, minimum height=0.5cm,thick]{};
\node (c2) at (1.5,0.1) [circle, draw, minimum height=0.5cm,thick]{};
\node[isosceles triangle, draw, minimum size =.9cm] (T)at (2.4,-0.25){};
\node (d1) at (8.8,1.5) [rectangle, draw, right, minimum height=1.0cm]{\scriptsize ${D_1}$};
\node (d11) at (8.8,0.3) [rectangle, draw, right, minimum height=1.0cm]{\scriptsize ${D_2}$};
\node (d2) at (8.8,-1.5) [rectangle, draw, right, minimum height=1.0cm]{\scriptsize ${D_K}$};
\foreach \i in {-0.5,-0.6,-0.7} {\fill (9.15,\i) circle(0.02cm);}
\foreach \i in {-0.5,-0.6,-0.7} {\fill (4.95,\i) circle(0.02cm);}
\foreach \i in {-0.5,-0.6,-0.7} {\fill (5.95,\i) circle(0.02cm);}
\foreach \i in {-0.5,-0.6,-0.7} {\fill (6.9,\i) circle(0.02cm);}
\node () at (1.57,-0.57) {\small {SSMF}};
\node () at (2.55,-0.25) {\footnotesize {EDFA}};
%\node (e5) at (2.30,0) [rectangle, dashed, draw, minimum height=1.5cm, minimum width=2.10cm]{};
\node () at (-1.71,1.8) {\small $X_1^n$};
\node () at (-1.71,0.55) {\small $X_2^n$};
\node () at (-1.66,-1.8) {\small $X_K^n$};
\node () at (8.5,1.8) {\small $Y_1^n$};
\node () at (8.5,0.55) {\small $Y_2^n$};
\node () at (8.5,-1.8) {\small $Y_K^n$};
\node () at (3.5,2.5) {{Interference Channel \eqref{eq:Kuserapprox}}};
%\node () at (2.2,1.0) {\large $\times N_s$};
\node (distr) at (3.5,3) {$p(Y_1^n,\cdots,Y_K^n|X_1^n,\cdots,X_K^n)$};
%\node[draw, ellipse, minimum width=.4cm, minimum height=4cm] at (-1.25,0) (ell1) {};
%\node[draw, ellipse, minimum width=.4cm, minimum height=4cm] at (8.05,0) (ell2) {};
\node (ell1) at (3.5,0.5) [rectangle, draw, fill=yellow, opacity=0.2, minimum height=6.2cm, minimum width=9.2cm]{};
%\draw[-] (distr)--(distr-|ell1)--(ell1);
%\draw[-] (distr)--(distr-|ell2)--(ell2);
\draw[-,red] (t2.117) --++(-0.48,0) node[left]{};
\draw[-,blue] (t2.121) --++(-0.39,0) node[left]{};
\draw[-,orange] (t2.125) --++(-0.33,0) node[left]{};
\draw[-,green] (t2.129) --++(-0.40,0) node[left]{};
\draw[-,purple] (t2.133) --++(-0.50,0) node[left]{};
%\draw[<-] (t2.126) --++(-0.35,0) node[left]{};
\draw[<-,color=red] (e1) --++(-0.7,0) node[above]{\small$M_1$};
\draw[<-,color=blue] (e11) --++(-0.7,0) node[above]{\small$M_2$};
\draw[<-,color=purple] (e2) --++(-0.7,0) node[above]{\small$M_K$};
\draw[->,color=red] (d1) --++(0.7,0) node[above]{\small$\hat{M}_1$};
\draw[->,color=blue] (d11) --++(0.7,0) node[above]{\small$\hat{M}_2$};
\draw[->,color=purple] (d2) --++(0.7,0) node[above]{\small$\hat{M}_K$};
%\draw[->] (T.0)  -- (3.45,-0.27) node[above]{};
\draw[-,color=red] ($(T.-180)+(0,+0.08)$)  -- ($(T.180-|t1.90)+(0,+0.08)$);
\draw[-,color=blue] ($(T.-180)+(0,+0.04)$)  -- ($(T.180-|t1.90)+(0,+0.04)$);
\draw[-,color=orange] ($(T.-180)+(0,+0.00)$)  -- ($(T.180-|t1.90)+(0,+0.00)$);
\draw[-,color=green] ($(T.-180)+(0,-0.04)$)  -- ($(T.180-|t1.90)+(0,-0.04)$);
\draw[-,color=purple] ($(T.-180)+(0,-0.08)$)  -- ($(T.180-|t1.90)+(0,-0.08)$);
\draw[->,color=red] (e1) --++(1.55,0) node[above]{};
\draw[->,color=blue] (e11) --++(1.55,0) node[above]{};
\draw[->,color=purple] (e2) --++(1.55,0) node[above]{};
\draw[->,color=red] (e3.east) --++(0.44,0) node[above]{};
\draw[->,color=blue] (e33.east) --++(0.44,0) node[above]{};
\draw[->,color=purple] (e4.east) --++(0.44,0) node[above]{};
\draw[<-,color=red] (d1.west) --++(-1.05,0) node[midway, above]{};
\draw[<-,color=blue] (d11.west) --++(-1.05,0) node[midway, above]{};
\draw[<-,color=purple] (d2.west) --++(-1.05,0) node[midway, above]{};
\draw[<-,color=red] (e7.west) --++(-0.39,0) node[above]{};
\draw[<-,color=blue] (e77.west) --++(-0.39,0) node[above]{};
\draw[<-,color=purple] (e8.west) --++(-0.39,0) node[above]{};
\draw[->,color=red] (e7.east) --++(0.32,0) node[above]{};
\draw[->,color=blue] (e77.east) --++(0.32,0) node[above]{};
\draw[->,color=purple] (e8.east) --++(0.32,0) node[above]{};
\draw[->,color=red] (e9.east) --++(0.4,0) node[above]{};
\draw[->,color=blue] (e99.east) --++(0.4,0) node[above]{};
\draw[->,color=purple] (e10.east) --++(0.4,0) node[above]{};
\draw[-,color=red] (e5.east) --++(0.24,0) node[above](tt1){};
\draw[-,color=blue] (e55.east) --++(0.24,0) node[above](tt2){};
\draw[-,color=purple] (e6.east) --++(0.24,0) node[above](tt3){};
\coordinate (A) at (7.8,1.7);
\coordinate (B) at (7.55,1.5);
\draw[-,color=red] (B) -- (A) node[above]{};
\coordinate (C) at (7.8,0.5);
\coordinate (D) at (7.55,0.3);
\draw[-,color=blue] (D) -- (C) node[above]{};
\coordinate (E) at (7.8,-1.3);
\coordinate (F) at (7.55,-1.5);
\draw[-,color=purple] (F) -- (E) node[above]{};
\coordinate (AA) at (7.48,1.7);
\coordinate (BB) at (7.7,1.25);
\draw [<-,color=red] (BB) to [out=90,in=10] (AA);
\coordinate (CC) at (7.48,0.5);
\coordinate (DD) at (7.7,0.05);
\draw [<-,color=blue] (DD) to [out=90,in=10] (CC);
\coordinate (EE) at (7.48,-1.3);
\coordinate (FF) at (7.7,-1.75);
\draw [<-,color=purple] (FF) to [out=90,in=10] (EE);
%\draw [-] (e5) to[cspst] (d1);
\end{tikzpicture}}
%\vspace{0.35ex}
\caption{WDM interference channel model under consideration.}
\label{fig:ICmodel}
%\end{figure*}
\end{wrapfigure}
We study the system in Fig.~\ref{fig:ICmodel}, where the interference channel encompasses the electro-optical conversion, WDM multiplexing, channel, demultiplexing, receiver DSP, matched filtering and sampling.  
We assume single-polarization transmission and study a single span of standard single mode fiber (SSMF). A first-order regular perturbative discrete-time model~\cite{dar2013properties} can be used to express the output of user $k \in \{1,2,\ldots,K\}$:
\small
\begin{align} \label{eq:Kuser}
Y_{k}[i]&=X_{k}[i]+N_{k}[i]+j \gamma \sum_{t=-\infty}^{\infty} X_{k}[i-t] \sum_{l=-\infty}^{\infty} \sum_{m=-\infty}^{\infty} S_{k,w}^{l,m,t} \sum_{\substack{w=1\\w \neq k}}^K  X_{w}[i-l] X_{w}[i-m]^{*},
\end{align}
\normalsize
where $X_{k}[i]$ is the input of user$-k$ at instant $i \in \{1,2,\ldots,n\}$, $Y_{k}[i]$ is the output at time $i$, while $\gamma$ is the fiber nonlinearity parameter. The channel coefficients $S_{k,w}^{l,m,t}$ given in~\cite[eq. (7)]{dar2013properties} are computed numerically. The term $N_{k}[i]$ models additive complex Gaussian noise from the amplifier, assumed to be zero mean and variance $2\sigma_k^2$.
Length-$n$ codewords are assumed with peak power constraints, i.e., $
\max_{i \in \{1,2,\ldots,n\}} |x_{k}[i]|^2 \leq P_k, \: \forall k \in \{1,2,\ldots,K\}.
$

The largest contribution to the NLI for few-span systems of relatively short lengths utilizing lumped amplification comes from the $S_{k,w}^{m,m,0}$ terms in \eqref{eq:Kuser}, as observed in \cite[Figs. 4 and 5]{dar2016pulse} and \cite[eq. (8)]{dar2013properties}. We also truncate the sums in \eqref{eq:Kuser} to $-M\leq t,l,m\leq M$ to get the following approximate model:
\begin{align} \label{eq:Kuserapprox}
Y_{k}[i] &\approx 
X_{k}[i]\Biggl(1+j  \sum_{m=-M}^{M} \sum_{\substack{w=1\\w \neq k}}^K c_{k,w}^m |X_{w}[i-m]|^2\Biggr)+N_{k}[i],
\end{align}
 where $\gamma S_{k,w}^{m,m,0} \triangleq c_{k,w}^m$ for brevity.
In \cite{ramachandran2022wdm}, the XPM terms corresponding to $c_{k,w}^m$ were assumed to be independent of $w$, which we correct here, thus allowing for frequency separation-dependent coupling between the channels. 

Now let $\mathcal{C}_K$ denote the capacity region of the given system, i.e., the maximal set of jointly achievable rates $(R_1,R_2,\ldots,R_K)$ of all the users. The single-user capacity is $C_k \triangleq \max \{R_k| (R_1,R_2,\ldots,R_K) \in \mathcal{C}_K\}$. 
We now have the following theorem, whose proof is similar to the one given in \cite[App. A,B]{ramachandran2021wdm}. 
\vspace{-1mm}
\begin{theorem} \label{thm:OBgenieK}
The single-user capacity $C_k$ is bounded as
\small
\begin{align} \label{eq:capUB}
\log_2\Biggl(\!\!1+\frac{P_{k}}{2\sigma_k^2 e}\Biggl(\!\!1+\Biggl(\sum_{\substack{w=1\\w \neq k}}^K P_w \sum_{m=-M}^M c_{k,w}^m\!\Biggr)^2\Biggr)\!\!\! \Biggr) \leq C_k \leq \log_2\Biggl(\!\!1+\frac{P_{k}}{2\sigma_k^2}\Biggl(\!\!1+\Biggl(\sum_{\substack{w=1\\w \neq k}}^K P_w \sum_{m=-M}^M c_{k,w}^m\!\Biggr)^2\Biggr)\!\!\! \Biggr), \: \forall \: k \in \{1,2,\ldots,K\}.
\end{align}
\vspace{-2mm}
\end{theorem}

\begin{figure}[!t]
\captionsetup{justification=centering}
\vspace{-3cm}
\hspace{-1cm}
\begin{subfigure}[b]{0.73\linewidth}
\centering
\maxsizebox{2.0\linewidth}{.645\linewidth}{\input{fig2.tikz}} 
\vspace{-2mm}
\caption{Capacity bounds in Theorem~\ref{thm:OBgenieK} vs. input power}
\label{fig:OBsnr}
\end{subfigure}
\hspace{-20mm}
\begin{subfigure}[b]{0.35\textwidth}
\begin{tabular}{c c}
\toprule
Parameter & Value \\ %[0.5ex]
\midrule
Memory Length $M$ (eq. \eqref{eq:Kuserapprox}) & 5\\
Number of WDM users & 3 \\
Distance & $250 \: \textrm{km}$ \\
Nonlinearity parameter $\gamma$ & $1.2 \: \textrm{W}^{-1} \textrm{km}^{-1}$ \\
Signalling Rate & $32 \: \textrm{Gbaud}$ \\
Fiber attenuation $\alpha$ & $0.2 \: \textrm{dB/km}$ \\
Group velocity dispersion $\beta_2$ & $-21.7 \: \textrm{ps\textsuperscript{2}/km}$ \\
RRC pulse-shaping roll-off & 0.1\\
\bottomrule
\begin{minipage}{.1cm}
\vfill
\end{minipage}
\end{tabular}
%}
%\end{table}
\vspace{-2mm}
\caption{Model Parameters}
\label{table:param}
%\vspace{-.45cm}
\end{subfigure}
%\vspace{-2mm}
\caption{Bounds versus peak input power, and model parameters}
\end{figure}
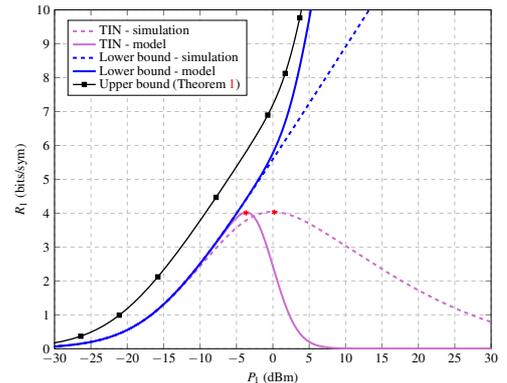
\begin{wrapfigure}{!t}{0.4\textwidth}
\vspace{-10mm}
\scalebox{0.5}{% This file was created by matlab2tikz.
%
%The latest updates can be retrieved from
%  http://www.mathworks.com/matlabcentral/fileexchange/22022-matlab2tikz-matlab2tikz
%where you can also make suggestions and rate matlab2tikz.
%
\definecolor{mycolor4}{rgb}{0.80000,0.42353,0.80000}%
\begin{tikzpicture}

\begin{axis}[%
width=4.521in,
height=3.538in,
at={(0.758in,0.509in)},
scale only axis,
xmin=-30,
xmax=30,
xlabel={$P_1~\text{(dBm)}$},
ymin=0,
ymax=10,
ylabel={$R_1$ (bits/sym)},
axis background/.style={fill=white},
xmajorgrids,
ymajorgrids,
legend style={
legend pos=north west,legend cell align=left,row sep=-0.5ex,grid style={dashed},font=\normalsize
}
]
\addplot [color=mycolor4, dashed, line width=1.5pt]
  table[row sep=crcr]{%
-35	0.0212086355661185\\
-32.2	0.0401461745729819\\
-30.1	0.0645556236496247\\
-28.4	0.0944857404379604\\
-26.9	0.131721205916328\\
-25.6	0.174978032299528\\
-24.5	0.221683708982759\\
-23.5	0.273888958504919\\
-22.5	0.337001466153133\\
-21.6	0.40448681788358\\
-20.7	0.483331306828539\\
-19.9	0.563861698942453\\
-19.1	0.654952981263321\\
-18.3	0.757170369853519\\
-17.5	0.870905784109667\\
-16.7	0.996340978559076\\
-15.9	1.13341877316731\\
-15.1	1.28182445477435\\
-14.3	1.4409776399577\\
-13.4	1.63180882883277\\
-12.5	1.83357026186221\\
-11.4	2.09206548760103\\
-10	2.43319615121745\\
-7.8	2.96990525099092\\
-6.9	3.17864600005078\\
-6.1	3.35348256416747\\
-5.4	3.49539759226373\\
-4.8	3.6069960671375\\
-4.2	3.707873666541\\
-3.6	3.79679097764901\\
-3.1	3.86099427298081\\
-2.6	3.91569893696066\\
-2.1	3.96058641248904\\
-1.6	3.99549888574864\\
-1.1	4.02044434753795\\
-0.600000000000001	4.0355924067456\\
-0.100000000000001	4.04126130916546\\
0.5	4.03618787101106\\
1.1	4.01907516061369\\
1.7	3.99102940903465\\
2.3	3.95323287837497\\
3	3.89840988853168\\
3.8	3.82385482634707\\
4.7	3.72781222265998\\
5.7	3.60963870496543\\
6.9	3.456673439205\\
8.5	3.24128470710036\\
15.8	2.24389175233177\\
17.7	2.00201704381283\\
19.4	1.79598560744907\\
21	1.61215769664953\\
22.6	1.43896564482982\\
24.1	1.28686116672841\\
25.6	1.14511603157905\\
27.1	1.01399578091432\\
28.6	0.893606290130577\\
30.1	0.783890885143684\\
31.6	0.684635562433151\\
33.2	0.589886818003272\\
34.8	0.506041939798422\\
36.5	0.428104421265822\\
38.2	0.360708040659404\\
40	0.299701164392495\\
};
\addlegendentry{TIN - simulation}

\addplot [color=mycolor4, line width=1.5pt]
  table[row sep=crcr]{%
-35	0.0212088757032944\\
-32.2	0.0401473623923465\\
-30.1	0.0645595355527604\\
-28.4	0.0944959346975764\\
-26.9	0.131744762462482\\
-25.6	0.175026342609286\\
-24.5	0.221771796526355\\
-23.5	0.274040005382425\\
-22.5	0.337258497781001\\
-21.6	0.404898450408183\\
-20.7	0.483985263110675\\
-19.9	0.564841256846442\\
-19.1	0.656409229079536\\
-18.3	0.759318001265541\\
-17.5	0.874046575642645\\
-16.7	1.00089459042355\\
-15.9	1.13996276974234\\
-15.1	1.29114584560302\\
-14.3	1.45413864351546\\
-13.5	1.62845388445451\\
-12.7	1.81344803797966\\
-11.8	2.03336586680064\\
-10.9	2.26455920397164\\
-9.9	2.53290905512615\\
-8.8	2.83922400854946\\
-6.1	3.59938703119052\\
-5.6	3.72720115125461\\
-5.2	3.820382506903\\
-4.8	3.90167866483552\\
-4.5	3.95211504344391\\
-4.2	3.99095223970475\\
-3.9	4.01555835665688\\
-3.7	4.02266846538362\\
-3.5	4.0213593508387\\
-3.3	4.01082088263029\\
-3.1	3.99031531324847\\
-2.9	3.95921961889668\\
-2.7	3.91706700315824\\
-2.5	3.86358298067969\\
-2.3	3.79871151494019\\
-2.1	3.72262768928859\\
-1.9	3.635735220901\\
-1.7	3.53864937317674\\
-1.5	3.43216791801152\\
-1.2	3.25692254190964\\
-0.899999999999999	3.06627166802443\\
-0.5	2.79474175743983\\
0.100000000000001	2.36856646610032\\
0.700000000000003	1.94430940887094\\
1.1	1.6742496286161\\
1.4	1.48251316453008\\
1.7	1.30219061098087\\
2	1.13476916358479\\
2.3	0.981293598133384\\
2.6	0.842353358148614\\
2.9	0.718092096897749\\
3.2	0.608243379386295\\
3.5	0.512190569581229\\
3.8	0.429043942700872\\
4.1	0.357725294489491\\
4.4	0.297050257472023\\
4.7	0.245800629789443\\
5.1	0.190074301564053\\
5.5	0.146352633869689\\
5.9	0.112305999081968\\
6.4	0.0803641178273296\\
7	0.0535727249616329\\
7.7	0.0332560121998569\\
8.6	0.0179499885809236\\
9.9	0.00733802224863922\\
12.1	0.00160839997895579\\
18.5	1.93472597374011e-05\\
40	6.86384282744257e-12\\
};
\addlegendentry{TIN - model}

\addplot [color=blue, dashed, line width=1.5pt]
  table[row sep=crcr]{%
-35	0.0212118583383827\\
-32.2	0.0401551051979609\\
-30.1	0.0645752590218294\\
-28.4	0.0945236382113777\\
-26.9	0.131790085533247\\
-25.6	0.175095253376142\\
-24.5	0.221869343046386\\
-23.5	0.274172889730728\\
-22.5	0.337438153010162\\
-21.6	0.405132375726367\\
-20.7	0.484287449083048\\
-19.9	0.565217916475916\\
-19.1	0.656875266269019\\
-18.3	0.759890165333672\\
-17.5	0.874743435236134\\
-16.7	1.00173653125473\\
-15.9	1.14097212918415\\
-15.1	1.29234741303986\\
-14.3	1.45556099103241\\
-13.5	1.63013248775648\\
-12.7	1.81543220046714\\
-11.8	2.03578964578151\\
-10.9	2.26761480506121\\
-9.9	2.5371619889879\\
-8.9	2.81769149601595\\
-7.8	3.13706733398246\\
-6.6	3.49618391130252\\
-5.2	3.92645903403457\\
-3.6	4.42963922468867\\
-1.7	5.03867622345392\\
0.600000000000001	5.78735157057073\\
3.6	6.77582354796472\\
7.4	8.03982515728192\\
11.7	9.48162538884809\\
15.6	10.8000718340961\\
19	11.9603981168002\\
21.9	12.9609966738151\\
24.5	13.8693253064154\\
24.9	14.0101920024951\\
};
\addlegendentry{Lower bound - simulation}

\addplot [color=blue, line width=1.5pt]
  table[row sep=crcr]{%
-35	0.0212088757033939\\
-32.2	0.040147362393661\\
-30.1	0.0645595355618838\\
-28.4	0.0944959347415661\\
-26.9	0.131744762639094\\
-25.6	0.175026343199583\\
-24.5	0.221771798168014\\
-23.5	0.274040009549061\\
-22.5	0.337258508372599\\
-21.6	0.40489847496606\\
-20.7	0.483985320117796\\
-19.9	0.564841377463381\\
-19.1	0.656409484454542\\
-18.3	0.759318542218196\\
-17.5	0.874047721834962\\
-16.7	1.00089701911357\\
-15.9	1.13996791492578\\
-15.1	1.29115674110178\\
-14.3	1.45416170171541\\
-13.5	1.62850264467576\\
-12.7	1.81355105646094\\
-11.8	2.03360457047141\\
-10.9	2.26511157366706\\
-9.9	2.534309863961\\
-8.9	2.81454300452334\\
-7.8	3.13377677434966\\
-6.6	3.4933004489468\\
-5.3	3.8946054959454\\
-4.1	4.27620831878078\\
-3.1	4.60468624590121\\
-2.3	4.87766419307949\\
-1.6	5.12789683237467\\
-1	5.35488494877176\\
-0.5	5.5564348855283\\
-0.100000000000001	5.72837940126119\\
0.300000000000004	5.91230224344093\\
0.700000000000003	6.1108405806923\\
1	6.27107149285244\\
1.3	6.44241761879909\\
1.6	6.62612507988011\\
1.9	6.8233507137619\\
2.2	7.03508722619362\\
2.5	7.26208955218134\\
2.8	7.5048129314145\\
3.1	7.76337292221351\\
3.4	8.03753456320285\\
3.7	8.32673271352194\\
4	8.63011978397483\\
4.3	8.94663243470261\\
4.7	9.38697483579189\\
5.1	9.84549302556586\\
5.5	10.3192182418375\\
6	10.9286058557426\\
6.6	11.6791313389405\\
7.3	12.5734519529837\\
8.2	13.7421461620657\\
8.4	14.0037897774929\\
};
\addlegendentry{Lower bound - model}

\addplot [color=black, mark = square*, mark size=1.5, mark repeat=5, mark phase=3, line width=1.0pt]
  table[row sep=crcr]{%
-35	0.0569387479462549\\
-32.6	0.0975529306107248\\
-30.7	0.148414437185494\\
-29.1	0.209907971212729\\
-27.7	0.282366326001565\\
-26.4	0.36919247903829\\
-25.2	0.469305416521067\\
-24.1	0.580394656278436\\
-23.1	0.699101318643329\\
-22.1	0.835868639742763\\
-21.1	0.991441220483189\\
-20.1	1.16609362275528\\
-19.1	1.35960237846081\\
-18.1	1.57126885153212\\
-17	1.82374574241037\\
-15.8	2.12047318423836\\
-14.6	2.43670032384599\\
-13.2	2.8264172426248\\
-11.7	3.2640843988591\\
-9.90000000000001	3.81030471952304\\
-7.8	4.46863451569131\\
-5.5	5.2092756807648\\
-3.6	5.83903310887627\\
-2.3	6.28899420902385\\
-1.40000000000001	6.61975027142837\\
-0.700000000000003	6.89625279879925\\
-0.100000000000001	7.15377034091018\\
0.399999999999999	7.38843643787959\\
0.899999999999999	7.64687757718504\\
1.3	7.87460315446236\\
1.7	8.12416931327461\\
2.1	8.39823488900258\\
2.5	8.69886020802652\\
2.9	9.02722702348491\\
3.3	9.38346139975988\\
3.7	9.76661338976868\\
4.1	10.1747999797216\\
4.6	10.7163081767425\\
5.1	11.2872096385049\\
5.7	12.0030113247601\\
6.4	12.8693687651288\\
7.3	14.015994494523\\
8.5	15.5774505612393\\
10.5	18.2154870306882\\
15	24.1886570472966\\
};
\addlegendentry{Upper bound ($\textrm{Theorem}~\ref{thm:OBgenieK}$)}

    \addplot [color=black, line width=1.0pt, only marks, mark=asterisk, mark options={solid, red}]
  table[row sep=crcr]{%
-3.7	4.01\\
};

    \addplot [color=black, line width=1.0pt, only marks, mark=asterisk, mark options={solid, red}]
  table[row sep=crcr]{%
0.2	4.03\\
};

\end{axis}

\begin{axis}[%
width=5.833in,
height=4.375in,
at={(0in,0in)},
scale only axis,
xmin=0,
xmax=1,
ymin=0,
ymax=1,
axis line style={draw=none},
ticks=none,
axis x line*=bottom,
axis y line*=left
]
\end{axis}
\end{tikzpicture}%
}
\caption{Capacity lower bounds based on the model \eqref{eq:Kuserapprox} versus SSFM. 
}
\label{fig:OBregion}
\vspace{-.45cm}
\end{wrapfigure}

\vspace{-5mm}

\section{Numerical Results} \label{sec:num}
\vspace{-2mm}
In the table shown as Fig.~\ref{table:param}, we summarize the parameters used in our numerical results. For $P_1=P_2=P_3$, the bounds on user$-1$ in Theorem~\ref{thm:OBgenieK} are plotted versus the input power in Fig.~\ref{fig:OBsnr}. Any rate below the lower bound in \eqref{eq:capUB} is achievable, which gives the shaded blue area, while the upper bound in \eqref{eq:capUB} gives an inadmissible region (shaded gray). The TIN bound is achieved by treating \smash{$j X_{k}[i] \sum_{m=-M}^{M} \sum_{w \neq k} c_{k,w}^m |X_{w}[i-m]|^2$} in \eqref{eq:Kuserapprox} as Gaussian noise, which is shown as the shaded purple area. The achievability of the shaded orange area is unknown. Our lower bound outperforms TIN for higher powers, but coincides with TIN for $P_1 < -6$ dBm.

In Fig.~\ref{fig:OBregion}, we compare the capacity bounds in Theorem~\ref{thm:OBgenieK} \& TIN obtained from the model \eqref{eq:Kuserapprox} with SSFM simulations. The simulations are performed with Gaussian inputs for all the users with variance $P_1$ and mismatched decoding. It is observed that the power at which the TIN curves peak differs between the model and simulations. Furthermore, it is observed that beyond a certain power ($0$ dBm in Fig.~\ref{fig:OBregion}), the capacity lower bound from simulations is smaller than that predicted by the model. This is chiefly due to the inaccuracy of the adopted model at higher powers.

\vspace{-1.8mm}

\section{Conclusions}
\vspace{-1.8mm}
%We presented achievable rates for an optical WDM system that outperform TIN, and validated it using SSFM simulations. 
Obtaining tighter lower bounds as well as designing schemes that can achieve the presented capacity bounds in practice, are among the avenues for further research.

\vspace{1mm}
\scriptsize
\noindent {\onehalfspacing The work of V.  Ramachandran, A. Barreiro and A. Alvarado has received funding from the ERC under the EU’s Horizon 2020 research and innovation programme via the Starting grant FUN-NOTCH (grant ID: 757791) . The work of G.~Liga is supported by the EuroTechPostdoc programme under the EU’s Horizon 2020 research and innovation programme (grant ID: 754462).}
\normalsize
\vspace{-2mm}
%\section{References}

\end{document}